\documentclass[11pt,draftcls,onecolumn]{IEEEtran}
\usepackage{cite}
\usepackage{amsmath,amssymb,amsfonts}
\usepackage{algorithmic}
\usepackage{amssymb}
\usepackage{algorithm}
\usepackage{graphicx}
\usepackage{textcomp}
\usepackage{hyperref}
\usepackage{caption}
\usepackage{xcolor}

\begin{document}
\title{Distributed Attribute-based Private\\ Access Control}
\author{Amir Masoud Jafarpisheh, Mahtab Mirmohseni, and Mohammad Ali Maddah-Ali\\
Department of Electrical Engineering, Sharif University of Technology, Tehran, Iran}

\maketitle

\newtheorem{Example}{\textbf{Example}}
\newtheorem{Remark}{Remark}
\newtheorem{Lemma}{Lemma}
\newtheorem{Claim}{Claim}
\newtheorem{Theorem}{\textbf{Theorem}}
\newtheorem{Definition}{\textbf{Definition}}

\begin{abstract}
In attribute-based access control, users with certain verified attributes will gain access to some particular data. Concerning with privacy of the users' attributes, we study the problem of distributed attribute-based private access control (DAPAC) with multiple authorities, where each authority will learn and verify only one of the attributes. 

To investigate its fundamental limits, we introduce an information theoretic DAPAC framework, with $N \in \mathbb{N}$, $N\geq 2$, replicated non-colluding servers (authorities) and some users. Each user has an attribute vector $\mathbf{v^*}=(v_1^*, ..., v_N^*)$ of dimension $N$ and is eligible to retrieve a message $W^{\mathbf{v}^*}$, available in all servers.  Each server $n\in [N]$ is able to only observe and verify the $n$'th attribute of a user. In response, it sends a function of its data to the user. The system must satisfy the following conditions: (1) \emph{Correctness:} the user with attribute vector $\mathbf{v^*}$ is able to retrieve his intended message  $W^{\mathbf{v}^*}$ from the servers' response, (2) \emph{Data Secrecy:}  the user will not learn anything about the other messages, (3) \emph{Attribute Privacy:} each Server~$n$ learns nothing beyond attribute $n$ of the user.  The capacity of the DAPAC is defined as the ratio of the file size and the aggregated size of the responses, maximized over all feasible schemes. We obtain a lower bound on the capacity of this problem by proposing an achievable algorithm with rate $\frac{1}{2K}$, where $K$ is the size of the alphabet of each attribute.
\end{abstract}

\section{Introduction}
In outsourcing the data storage to cloud servers, a mechanism, known as  access control, is required to guarantee users' access to the appropriate data. For individual use, access control can be simply designed using asymmetric cryptography. 
Data, encrypted by a public key,  can be decrypted by the user with the corresponding private key. However, in some cases, data recipient is not known at the encryption time or the data is intended for a group of users.
For example, the users of an industrial cloud may include sales enterprises, consulting firms, manufacturing enterprises, logistics enterprises, and scientific research institutions, where each group may be granted to access some specific data~\cite{song2019efficient}.
Attribute-based access control can be considered as a solution for these circumstances, where only users whose attributes satisfy a pre-specified access policy can access the data.

In an attribute-based access control, a user with a certain set of attributes is eligible to gain access to specific data.
In a central solution, one authority is responsible to verify users' attributes. The concern is the central authority will learn all the attributes of the users, which raises serious privacy issues.
For example, consider the case, where a patient with certain range of income and with a particular disease is eligible to have access to some information. However, for various reasons, he does not want a central authority to know both of his income and his disease. This concern can be resolved by delegating the task of attribute verification to multiple authorities. For example, one authority, say financial organizations, \emph{only} observes and verifies the income attribute of the patient, and another authority \emph{only} observes and verifies the disease attribute, without learning anything about another attribute of the patient. The access is granted if both attributes have been verified.  

There are algorithms for non-centralized attribute-based access control based on cryptographic primitives, e.g., bilinear mappings, hash functions, and encryption algorithms~\cite{wang2011hierarchical,jung2013privacy}. In this paper, we propose an information theoretic framework for the problem of distributed attribute-based private access control (DAPAC), and investigate its fundamental limits.

\textbf{\textit{Related Works}:} The idea of identity-based encryption was first proposed by Shamir~\cite{shamir1984identity}. In~\cite{boneh2001identity}, the first fully functional identity-based encryption scheme was introduced. In~\cite{sahai2005fuzzy}, attribute-based encryption systems are introduced as a special case of identity-based encryption systems, where each user is specified by an attribute vector. 
Among its many applications, attribute-based access control is proposed for the personal health record services~\cite{qian2015privacy, zhang2019hidden}. In the scheme of~\cite{sahai2005fuzzy}, it is assumed that there is an authority that verifies all attributes of the user. The systems with multiple authorities to verify the attributes is studied in~\cite{chase2007multi,jung2013privacy}. 

To the best of our knowledge, all existing works on the attribute-based access control problem utilize cryptographic primitives. In contrary, in this work, we take an information theoretic approach. Our work is mainly inspired by the results on information theoretic private information retrieval (PIR) ~\cite{sun2017capacity, sun2018capacity, banawan2019private, banawan2020capacity, ulukus2022private}. Here, we elaborate on the similarities and the differences between the PIR and DAPAC problems. In terms of similarities, in both problems:

(i) The user tends to retrieve a message from some replicated servers,

(ii) The user wishes to keep some information about the index of the desired message private from each server. 

(iii) In DAPAC, the user should gain no information about the non-requested messages, as in symmetric PIR~\cite{sun2018capacity}.

Despite the above similarities, there are some intrinsic differences between these two problems:

(i) In PIR, the index of the requested message is kept entirely private from all the servers. However, in DAPAC, the index of message is an attribute vector, and each server is supposed to observe and verify one of the attributes, without learning any information about other attributes.

(ii) In PIR, all the files in each server is basically accessible for the user. Of course, the one that the user is asked for is revealed to the user, following the protocol. However, in DAPAC, when server $n$ verifies the $n$'th attribute of the user, in that server, only the messages with index vectors with the matched $n$'th entry will be accessible to the user. Thus, after attribute verification at each server, the content of the servers are not replicated from user's perspective. Because of these differences, the solutions of the PIR problem are not applicable to the DAPAC problem.

\textbf{\textit{Our Contribution}:} We propose an information theoretic framework for the DAPAC problem. In the proposed model, there is a user with $N$ attributes, denoted by attribute vector $\mathbf{v}^*=(v_1^*, ..., v_N^*)$, where each attribute has $K$ possible values. The user have the right (and wishes) to access the message $W^{\mathbf{v}^*}$, with access policy $\mathbf{v}^*$. 
There are $N$ replicated servers, containing all messages,  where Server $n$ can verify $v_n^*$ and in response, it will release a function of its content.  
We consider the access control and privacy constraints. The access control constraint assures that the user is able to retrieve his intended message $W^{\mathbf{v}^*}$ from what it receives from the servers (correctness), and he gains no information about other messages (data secrecy). The (user's attribute) privacy constraint guarantees that each server gains no information about the other attributes of the user, except the one for which that server is responsible for its verification. 
The goal is to minimize the download cost.  The capacity of the DAPAC is defined as the ratio of the file size and the aggregated size of the responses, maximized over all feasible schemes. We obtain a lower bound on the capacity of this problem by proposing an achievable algorithm.

In the proposed algorithm, the user proves his $n$-th attribute to Server~$n$ and after verifying $v_n^*$, Server~$n$ authorizes the user's access to the message set $\mathcal{W}^{v^*_n}$. The user to retrieve the desired message $W^{\mathbf{v}^*}\in \mathcal{W}^{v^*_n}$ sends queries to Server~$n$ in the form of linear combinations of messages that have two attributes in common; Obviously, one of them is $v_n^*$. Considering any two servers,
the user downloads two linear combinations of messages with the same access policy and the same message indices, e.g., $L_m$ and $L_n$ from Servers $m$ and $n$, respectively. Although the servers add an independent part of common randomness to each of the requested linear combinations to guarantee the data secrecy, the added randomness is the same for the linear combinations $L_m$ and $L_n$. Therefore, the user can subtract these two linear combinations to retrieve a chunk of the desired message $W^{\mathbf{v}^*}$.
The user can make ${N \choose 2}$ such linear combinations to completely retrieve the message $W^{\mathbf{v^*}}$. In this scheme, to guarantee privacy, at each Server~$n\in [N]$,  the distribution on the other attributes of the user (by observing the queries) is uniform, because all the linear combinations of messages that have two attributes in common (one is $v_n^*$) are requested. So each server learns nothing about the other attributes of the user.

The rest of the paper is as follows. Section~\ref{System_model} formally introduces our proposed information theoretic framework. Section~\ref{Main_results} presents main results. Section~\ref{achievable} presents the achievable algorithm, and Section~\ref{proof} contains the proofs.

\section {System Model}
\label{System_model}
\begin{figure}[tb]
\centering
\includegraphics[trim={6cm .1cm .5cm 0cm},clip,scale=.9]{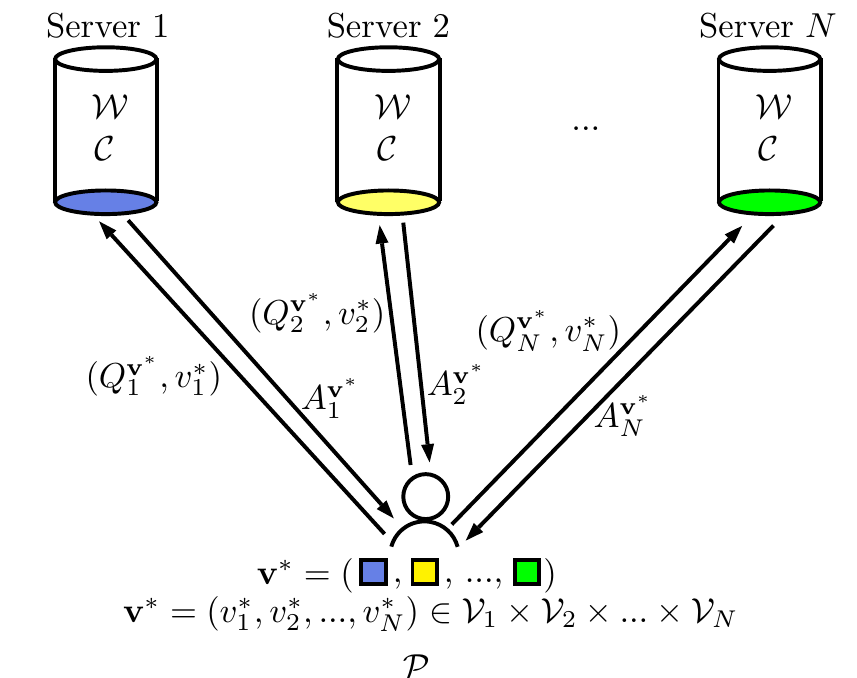}
\captionsetup{justification=centering}
\caption{System model of the DAPAC}
\label{system_model}
\end{figure}

As shown in Fig.~\ref{system_model}, we consider a system, including $N\geq 2$ non-colluding semi-honest servers, each storing an identical copy of a database of messages $\mathcal{W}$ and a set of common randomness $\mathcal{C}$, and a user with $N$ attributes, shown by attribute vector $\mathbf{v^*}=(v^*_1, ..., v^*_N)$, who wishes to download a message from the servers that corresponds to his attribute vector.
Each server is responsible for verifying one of the attributes, i.e., Server $n$ is responsible for verifying $v_n^*$. The user can show the evidence of possessing attribute $v^*_n$ to Server $n$ and he cannot falsify the possessing of any other attribute $v_n$, $\forall v_n\neq v_n^*$ and $v_n \in \mathcal{V}_n$. 
There are $N$ disjoint attribute sets $\mathcal{V}_1, ..., \mathcal{V}_N$ and $|\mathcal{V}_n|=K\geq 2$ for $n\in [N]$. So, $v_n$, attribute $n$, can take one of $K\geq 2$ values from the set $\mathcal{V}_n$. Moreover, the user has access to an independent uniform permutation $\mathcal{P}$.

We define an access policy for each message such that if the access policy of a message is $(v_1, ..., v_N)$, then this message is shown as $W^{(v_1, ..., v_N)}$, and only users with attribute vector $(v_1, ..., v_N)$ have the right to access it. All messages have equal length, so for $v_n \in \mathcal{V}_n$, $n \in [N]$, we have
\begin{equation}
H(W^{(v_1, ..., v_N)})=L.
\end{equation}

The messages of different access policies are independent. Let $\mathcal{V}^N\doteq\mathcal{V}_1\times \mathcal{V}_2\times...\times \mathcal{V}_N$, so for each $\mathcal{V}\subset\mathcal{V}^N$, and $\Tilde{\mathcal{W}}=\{W^{(v_1, ..., v_N)}, (v_1, ..., v_N)\in \mathcal{V}\}$, we have
\begin{align}
H(\Tilde{\mathcal{W}})=\sum_{(v_1, ..., v_N)\in \mathcal{V}}H(W^{(v_1, ..., v_N)}).
\end{align}

The user can send queries for messages with different access polices $\mathbf{v}^{(n)}$ for $\forall n \in[N]$, and
\begin{align}
 \mathbf{v}^{(n)}=(v_1^{(n)}, ..., v_n^*, ..., v_N^{(n)})\in \mathcal{V}^N.  
\end{align}

The user sends query $Q_n^{\mathbf{v}^{(n)}}$, and his $n$-th attribute, $v^*_n$, to Server~\textit{n} as a pair $(Q_n^{\mathbf{v}^{(n)}}, v_n^*)$. The server verifies the attribute $n$ of the user correctly (by verifying the possession evidence) and let the user access the messages $\mathcal{W}^{v^*_n}=\{W^{(v_1,\ldots,v_n,\ldots,v_N)}\in\mathcal{W}|v_n=v^*_n\}$, if the user is verified to have attribute $v_n^*$. When a user wants to retrieve his corresponding message labelled with $\mathbf{v^*}=(v^*_1, ..., v^*_N)\in \mathcal{V}^N$, he sends query pairs $(Q_n^{\mathbf{v}^*}, v_n^*)$ to Server $n$. The Server verifies the attribute $n$ of the user correctly and let the user access the messages $\mathcal{W}^{v^*_n}$, if the user is verified to have attribute $v_n^*$.

The queries are generated with no knowledge about the messages. So, for each $\mathbf{v}=(v_1, ..., v_N)\in \mathcal{V}^N$, we have
\begin{equation}
I(Q_1^{\mathbf{v}}, Q_2^{\mathbf{v}}, ..., Q_N^{\mathbf{v}}; \mathcal{W})=0,
\end{equation}
where $Q_n^{\mathbf{v}}$ is the query sent to Server $n$ to access the message with access policy $\mathbf{v}$.
The queries are deterministic functions of the user's attributes and the randomness $\mathcal{P}$, used by the user to generate queries. So, for each $\mathbf{v}=(v_1, ..., v_N)\in \mathcal{V}^N$,
\begin{equation}
H(Q_1^{\mathbf{v}}, Q_2^{\mathbf{v}}, ..., Q_N^{\mathbf{v}}|\mathcal{P}, v_1, v_2, ..., v_N)=0.
\end{equation}

The Server~\textit{n}, after receiving the query for the message  ${W}^{(\mathit{v}_1,\ldots, \mathit{v}^*_n, \ldots, \mathit{v}_N)}$ (i.e., $Q_n^{\mathbf{v}}$, $\mathbf{v}=(v_1, \ldots, v_n^*, \ldots, v_N)$), generates the answer set $A_n^{\mathbf{v}}$ based on the received query, attribute $n$ of the user, $v_n^*$, the messages that correspond to the attribute $v_n^*$ (i.e., $\mathcal{W}^{v_n^*}$), and the common randomness between servers, i.e., $\mathcal{C}$. So, for each $n\in[N]$,
\begin{equation}
H(A^{\mathbf{v}}_n|Q^{\mathbf{v}}_n, v^*_n, \mathcal{W}^{{v}^*_n}, \mathcal{C})=0.
\end{equation}

Now, we define the constraints to guarantee the access control and the privacy of the other attributes of the user.

\textbf{Access control}: To ensure the access control in our setup, each user must correctly retrieve his own message (correctness), while preventing the leakage about the other messages to him (data secrecy). Hence, it is required that for $\mathcal{A}:= \{A_1^{\mathbf{v}^{(1)}}, ..., A_N^{\mathbf{v}^{(N)}}\}$, and $\mathcal{Q}:= \{ (Q_1^{\mathbf{v}^{(1)}}, v^*_1), ..., (Q_N^{\mathbf{v}^{(N)}}, v^*_N)\}$:

(i) The user can retrieve his message:
\begin{align}[Correctness]
\label{correctness}
\:\:H(W^{\mathbf{v^*}}|\mathcal{A} ,\mathcal{Q},\mathcal{P})=0.
\end{align}

(ii) Secrecy of the other messages is preserved: 
\begin{align}[Data\text{ }Secrecy] 
\label{data_sec_com}
\:\:I(\mathcal{W}\backslash W^{\mathbf{v}^*}; \mathcal{A},\mathcal{Q},\mathcal{P}|{W}^{\mathbf{v}^*})=0.
\end{align}

\textbf{Privacy}: To preserve the user's privacy, it is required that the attribute vector be kept hidden from each Server~$n$ except $v^*_n$. Thus,

(iii) In Server $n$, for $n\in[N]$:
\begin{align}[Privacy]
\label{poa}
\:\:H(\{v^*_i: i\in[N], i\neq n\}|Q_n^{\mathbf{v}^{(n)}}, \mathcal{C}, \mathcal{W}, v^*_n)
=H(\{v^*_i: i\in[N], i\neq n\}| \mathcal{C}, \mathcal{W}, v^*_n).
\end{align}

An $(N, K)$ DAPAC scheme for the above setup and for a set of vectors $\{\mathbf{v}^{(1)},\ldots,\mathbf{v}^{(N)}\}$ consists of query-answer functions $(Q_n^{\mathbf{v}^{(n)}},A_n^{\mathbf{v}^{(n)}})$ for $n\in[N]$, and the corresponding decoding functions that map them to $W^{\mathbf{v^*}}$, common randomness $\mathcal{C}$, and random permutation $\mathcal{P}$. The retrieval rate of this code is the ratio of bits of the desired message (\textit{L}) to the total download cost from all servers in bits, i.e., $D=\sum_{n=1}^{N}H(A_n^{\mathbf{v}^{(n)}})$ and is defined as,
\begin{align}
R:=\frac{L}{D}.
\label{eq_r0}
\end{align}

\begin{Definition}
A rate $R$ is achievable if a DAPAC scheme with the retrieval rate greater than or equal to $R$ exists that satisfies the constraints of correctness \eqref{correctness}, secrecy of other messages \eqref{data_sec_com}, and privacy \eqref{poa} for all $\mathbf{v}^*=(v_1^*, v_2^*, ..., v_N^*)\in \mathcal{V}^N$. The capacity of the DAPAC problem is defined as,
\begin{align}
C:=\sup\{R:\text{ }R\text{ }\text{is achievable}\}.    
\end{align}

\end{Definition}

We also define a parameter to show the number of equations downloaded from all servers for the desired message retrieval. The download complexity is defined as,
\begin{align}
\label{download_comp}
DC:=\sum_{n=1}^{N}|A_n^{\mathbf{v}^{(n)}}|.
\end{align}

\section{Main Results}
\label{Main_results}
The first theorem presents a lower bound on the capacity of the DAPAC problem, and the second theorem presents the minimum common randomness required in Theorem~\ref{thoerem_1}.

\begin{Theorem}\label{thoerem_1}
In an $(N, K)$ DAPAC system, with at least two attributes ($N\geq 2$), where each has at least two values ($K=|\mathcal{V}_n|\geq 2$), the following rate is achievable with download complexity of $O(KN^2)$,
\begin{align}
\label{r_poly}
R=\frac{1}{2K}\leq C.
\end{align}
\end{Theorem}
\begin{IEEEproof}
To prove this lower bound, we propose an achievable algorithm with the rate $R=\frac{1}{2K}$ in Section~\ref{achievable}, and the rest of proof is provided in Section~\ref{theorem_1_proof}.
\end{IEEEproof}

\begin{Remark}
Consider an $(N, K)$ DAPAC system, we can run an $(N-1,N)$ secret sharing on the messages with different access policies, and then download all accessible messages from each server. This is a naive solution that jointly satisfies \eqref{correctness}, \eqref{data_sec_com}, and \eqref{poa}. In this scheme, the download complexity is $NK^{N-1}$ and
the size of each secret share is $L$ bits. So the achievable rate is $R_\mathsf{Naive}=\frac{1}{NK^{N-1}}$, and thus,
\begin{align}
\frac{R}{R_\mathsf{Naive}}=\frac{NK^{N-1}}{2K}=\frac{NK^{N-2}}{2}.
\end{align}

We observe that:

(i) For a fixed $K$, the proposed DAPAC scheme has an exponential gain over the naive scheme as $N$ increases.

(ii) For a fixed $N$, the proposed DAPAC scheme has a polynomial gain over the naive scheme as $K$ increases.

(iii) The download complexity of the proposed DAPAC scheme is less than the naive scheme.
\end{Remark}
\begin{Remark}
To guarantee the privacy, the user should hide the value of his other attributes in all their possible values (from each server). The larger the alphabet of attribute be, the harder is to provide the privacy. This result is reflected from \eqref{r_poly} as the achievable rate decreases when $K$ increases.
\end{Remark}
\begin{Remark}
The surprising fact about the achievable rate \eqref{r_poly} is that it is independent of the number of attributes $N$. The reason is that, in our achievable scheme, we split the messages into $\frac{N(N-1)}{2}$ equal chunks, each with length $\frac{L}{\frac{N(N-1)}{2}}$ bits. Then, we download $KN(N-1)$ linear combinations of these chunks to retrieve the desired message. So, the total download is $2KL$ bits, which is independent of $N$. 
\end{Remark}

To guarantee the data secrecy constraint \eqref{data_sec_com}, we need a minimum amount of independent common randomness $\mathcal{C}$ between servers. The following theorem presents the minimum common randomness required in the proposed achievable scheme, and its proof is provided in Subsection~\ref{theorem_2_proof}.
\begin{Theorem}
\label{thoerem_2}
In the proposed $(N, K)$ DAPAC scheme, the lower bound on the amount of common randomness is as,
\begin{align}
H(\mathcal{C})\geq K^2L.
\end{align}
\end{Theorem}
\section{Achievable algorithm}
\label{achievable}
In this section, we first present the key ideas of our proposed achievable scheme by a motivating example. Then we present the general achievable algorithms.

\textit{\textbf{Motivating Example:}}
Consider a $(3, 2)$ DAPAC system. Let $\mathcal {V}_1=\{\mathsf{M}, \mathsf{P}\}$, where $\mathsf{M}$ and $\mathsf{P}$ indicate the MSc and PhD, respectively, $\mathcal {V}_2=\{\mathsf{E}, \mathsf{C}\}$ where $\mathsf{E}$  and $\mathsf{C}$ indicate the Electrical Engineering and Computer Science, respectively, and $\mathcal {V}_3=\{\mathsf{S}, \mathsf{F}\}$ where $\mathsf{S}$ and $\mathsf{F}$ indicate Spring intake and Fall intake, respectively.
$\forall (v_1, v_2, v_3) \in \mathcal{V}^3$, we split the message $W^{(v_1, v_2, v_3)}$ into three equal chunks as $W^{(v_1,v_2,v_3)}=w_1^{v_1v_2v_3}||w_2^{v_1v_2v_3}||w_3^{v_1v_2v_3}$.
There are three servers; Each is responsible for verifying one of the attributes and giving messages to the user based on his attribute and requests. Suppose a user who needs to access the message $W^{(\mathit{\mathsf{M}, \mathsf{E}, \mathsf{S}})}$. The user commits $v^*_1=\mathsf{M}$, $v^*_2=\mathsf{E}$, and $v^*_3=\mathsf{S}$, in servers $1$, $2$, and $3$, respectively. Using a uniform random permutation $\mathcal{P}$, the user permutes the index of different chunks of messages and accesses the couple of messages as shown in Table~\ref{access_table}.
\begin{table}[tb]
\centering
\caption{Access table for the motivating example}
\label{access_table}
\scalebox{1}{
\begin{tabular}{|c|c|c|}
 \hline
$S\mathit{\{\mathsf{M}, \mathsf{P}\}}$  & $S\mathit{\{\mathsf{E}, \mathsf{C}\}}$ & $S\mathit{\{\mathsf{F}, \mathsf{S}\}}$ \\ \hline
 $\mathbf{w}_{\mathsf{M}}^{(1),1}=(w^{\mathsf{MES}}_{1},w^{\mathsf{MCS}}_{1})$ &$\mathbf{w}_{\mathsf{E}}^{(2),1}=(w^{\mathsf{MES}}_{2},w^{\mathsf{MEF}}_{1})$ & $\mathbf{w}_{\mathsf{S}}^{(3),1}=(w^{\mathsf{MES}}_{1},w^{\mathsf{MCS}}_{1})$ \\ 
$\mathbf{w}_{\mathsf{M}}^{(1),2}=(w^{\mathsf{MES}}_{2},w^{\mathsf{MEF}}_{1})$ &$\mathbf{w}_{\mathsf{E}}^{(2),2}=(w^{\mathsf{PES}}_{1}, w^{\mathsf{MES}}_{3})$ & $\mathbf{w}_{\mathsf{S}}^{(3),2}=(w^{\mathsf{PES}}_{1}, w^{\mathsf{MES}}_{3})$ \\ 
$\mathbf{w}_{\mathsf{M}}^{(1),3}=(w^{\mathsf{MCS}}_{2},w^{\mathsf{MCF}}_{1})$ &$\mathbf{w}_{\mathsf{E}}^{(2),3}=(w^{\mathsf{PES}}_{2},w^{\mathsf{PEF}}_{1})$ & $\mathbf{w}_{\mathsf{S}}^{(3),3}=(w^{\mathsf{MCS}}_{3},w^{\mathsf{PCS}}_{1})$ \\ 
$\mathbf{w}_{\mathsf{M}}^{(1),4}=(w^{\mathsf{MCF}}_{2},w^{\mathsf{MEF}}_{2})$ &$\mathbf{w}_{\mathsf{E}}^{(2),4}=(w^{\mathsf{MEF}}_{3},w^{\mathsf{PEF}}_{2})$ & $\mathbf{w}_{\mathsf{S}}^{(3),4}=(w^{\mathsf{PCS}}_{2},w^{\mathsf{PES}}_{3})$ \\ \hline
\end{tabular}}
\end{table}

The user generates $12$ vectors $\mathbf{a}^{(n)}_i$ for $ i \in [4], n \in [3]$, each a $1\times 2$ binary vector; Nine of them have random elements with independent uniform distribution in $\{0,1\}$ and the rest are:
\begin{align}
\mathbf{a}^{(2)}_1&=\mathbf{a}^{(1)}_2\oplus(1,0),\\
\mathbf{a}^{(3)}_1&=\mathbf{a}^{(1)}_1\oplus(1,0),\\
\mathbf{a}^{(3)}_2&=\mathbf{a}^{(2)}_2\oplus(0,1).
\end{align}

Then, the user sends queries for message $W^{(\mathit{\mathsf{M}, \mathsf{E}, \mathsf{S}})}$ as shown in Table \ref{table}, where $\forall i\in[9]$, ${s}_{i}$ is an independent part of $\mathcal{C}$.
\begin{table}[tb]
\centering
\caption{Request table for the motivating example} 
\label{table}
\scalebox{1}{
\begin{tabular}{|c|c|c|}
 \hline
$S\mathit{\{\mathsf{M}, \mathsf{P}\}}$  & $S\mathit{\{\mathsf{E}, \mathsf{C}\}}$ & $S\mathit{\{\mathsf{F}, \mathsf{S}\}}$ \\ \hline
 $\mathbf{a}^{(1)}_{1}.\mathbf{w}_{\mathsf{M}}^{(1),1}+s_1$ &$\mathbf{a}^{(2)}_{1}.\mathbf{w}_{\mathsf{E}}^{(2),1}+s_2$ & $\mathbf{a}^{(3)}_{1}.\mathbf{w}_{\mathsf{S}}^{(3),1}+s_1$ \\ 
$\mathbf{a}^{(1)}_{2}.\mathbf{w}_{\mathsf{M}}^{(1),2}+s_2$ &$\mathbf{a}^{(2)}_{2}.\mathbf{w}_{\mathsf{E}}^{(2),2}+s_3$ & $\mathbf{a}^{(3)}_{2}.\mathbf{w}_{\mathsf{S}}^{(3),2}+s_3$ \\ 
$\mathbf{a}^{(1)}_{3}.\mathbf{w}_{\mathsf{M}}^{(1),3}+s_4$ &$\mathbf{a}^{(2)}_{3}.\mathbf{w}_{\mathsf{E}}^{(2),3}+s_6$ & $\mathbf{a}^{(3)}_{3}.\mathbf{w}_{\mathsf{S}}^{(3),3}+s_8$ \\ 
$\mathbf{a}^{(1)}_{4}.\mathbf{w}_{\mathsf{M}}^{(1),4}+s_5$ &$\mathbf{a}^{(2)}_{4}.\mathbf{w}_{\mathsf{E}}^{(2),4}+s_7$ & $\mathbf{a}^{(3)}_{4}.\mathbf{w}_{\mathsf{S}}^{(3),4}+s_9$ \\ \hline
\end{tabular}}
\end{table}
Obtaining the corresponding answers, the user can retrieve the message $W^{(\mathit{\mathsf{M}, \mathsf{E}, \mathsf{S}})}=w^{\mathsf{MES}}_{1}||w^{\mathsf{MES}}_{2}||w^{\mathsf{MES}}_{3}$ correctly, because:
\begin{align}
w_1^{\mathsf{MES}}= \mathbf{a}^{(3)}_{1}.\mathbf{w}_{\mathsf{S}}^{(3),1}+s_1-(\mathbf{a}^{(1)}_{1}.\mathbf{w}_{\mathsf{M}}^{(1),1}+s_1),\\
w_2^{\mathsf{MES}}= \mathbf{a}^{(2)}_{1}.\mathbf{w}_{\mathsf{E}}^{(2),1}+s_2-(\mathbf{a}^{(1)}_{2}.\mathbf{w}_{\mathsf{M}}^{(1),2}+s_2),\\
w_3^{\mathsf{MES}}= \mathbf{a}^{(3)}_{2}.\mathbf{w}_{\mathsf{S}}^{(3),2}+s_3-(\mathbf{a}^{(2)}_{2}.\mathbf{w}_{\mathsf{E}}^{(2),2}+s_3).    
\end{align}

The user gains no information about the other messages, since we have used an independent part of common randomness $\mathcal{C}$, i.e., $s_i$, for each linear combination that includes the other messages. The user attributes privacy is also preserved. The reason follows. 
Server $1$ verifies $\mathsf{M}$, and the user wants to hide his other attributes from Server~1, i.e., $\mathsf{E}$ and $\mathsf{S}$.

(i) The structure of message vectors requested from Server $1$, $\{\mathbf{w}_{\mathsf{M}}^{(1),i}:\forall i \in [4]\}$, is independent of the second and the third attributes of the user, since it is composed of all message vectors that have two attributes in common and one of these common attributes is $\mathsf{M}$. 

(ii) The index of chunks of messages is determined by applying the random permutation $\mathcal{P}$ on the index of message chunks, so the index of messages reveals no information about the attributes of the user.

(iii) The elements of the coefficients vectors, $a_i^{(1)}$, $\forall i \in [4]$, have an independent uniform distribution on $\{0, 1\}$, and reveal no information about the attributes of the user. 

Therefore, the privacy of the user is preserved in Server $1$. A similar argument can be applied in Servers $2$ and $3$. The retrieval rate of DAPAC is $R=\frac{L}{12\times \frac{L}{3}}=\frac{1}{4}=\frac{1}{2K}$. 
\label{ex_poly_3}

To show the general achievable algorithm, we need to define the concept of \emph{type} of messages first.
\begin{Definition}
Consider $V_1$ as a vector of messages where  $J\geq 1$, and $\forall j \in [J]$, $\mathbf{v}^{(j)}$ is the access policy of each message and $i_j$s are the message indices: $${V_1}=(w^{\mathbf{v}^{(1)}}_{i_1}, ..., w^{\mathbf{v}^{(J)}}_{i_J}).$$
We define the type of messages in $V_1$, as a set $T(V_1)$, composing of the access policy of the messages in $V_1$. Thus,
\begin{align}
    T(V_1) = \{\mathbf{v}^{(1)}, \mathbf{v}^{(2)}, ..., \mathbf{v}^{(J)}\}.
\end{align}
 \end{Definition}
\begin{Definition}
Two vector of messages like $V_1$ and $V_2$ are of the same type if and only if, 
\begin{align}
T(V_1) = T(V_2).
\end{align}
\end{Definition}

\begin{Definition}
\label{def_4}
In Server $n$, $n \in [N]$, for each $k \in [K]$, first cast the set $\mathcal{V}_n$ into a list $\Tilde{\mathcal{V}_n}$:$\Tilde{\mathcal{V}_n} =$ List$(\mathcal{V}_n)$, then for $j\neq n$ define $U^{(n)}$, with entries
\begin{align}
U^{(n)}(k, j) := \{\mathbf{v} = (v_1, ..., v_N)\in \mathcal{V}^{N}| v_n = v_n^*, v_j = \mathcal{\Tilde{V}}_j(k)\}.    
\end{align}
\end{Definition}

\begin{algorithm}[b]
\caption{Initializing Algorithm for an $(N, K)$ DAPAC}
\label{alg_ini_poly}
\begin{algorithmic}[1]
\STATE Consider an $(N,K)$ DAPAC system with $N\geq2$, $K\geq 2$.\nolinebreak[4]
\STATE Split each message into $\frac{N(N-1)}{2}$ equal chunks.
\STATE For each subset of $K^{N-2}$ messages that have two attributes in common, assign an independent part of common randomness $\mathcal{C}$, e.g., $s_i$ with length $\frac{L}{\frac{N(N-1)}{2}}$. 
\end{algorithmic}
\end{algorithm}
The achievable DAPAC scheme comprises three algorithms; Initializing algorithm, user-side algorithm, and server-side algorithm. The initializing algorithm is run by the operator of the system or the servers themselves and is described in Algorithm \ref{alg_ini_poly}. When a user with attribute vector $\mathbf{v}^*$ wants to retrieve his related message, he sends queries to the $N$ servers and requests for linear combinations of messages to retrieve the message with access policy $\mathbf{v}^*$. The user-side algorithm is described in Algorithm~\ref{alg_us_poly}. The server receives a query, it verifies whether the user is an authorized user and whether only one linear combination from each type of messages exists in the query. If both hold, the server responds to the query based on the server-side algorithm, described in Algorithm \ref{alg_ss_poly}.

\begin{algorithm}[tb]
\caption{User Side Algorithm of an $(N, K)$ DAPAC}
\label{alg_us_poly}
\begin{algorithmic}[1]
\STATE For a user with attribute vector $\mathbf{v}^*=(v^*_1, v^*_2, ..., v^*_N)$:
\STATE  $Q^{\mathbf{v}^*}_n= \{\}$.
\STATE Permute the index of different messages chunks with a private and uniform random permutation $\mathcal{P}$.
\FOR {$n \in [N]$}
\STATE $\beta = 1$.
\FOR {$k \in [K]$}
\FOR {$j \in [N]\setminus n$}
\STATE  $\mathbf{w}_{v_n^*}^{(n),\beta}$ = A vector of messages with access policies given in  $U^{(n)}(k, j)$.
\IF {$\exists m \in [n-1]$ and $\exists \alpha \in [K(N-1)]$ such that $T(\mathbf{w}_{v_n^*}^{(n),\beta}) == T(\mathbf{w}_{v_m^*}^{(m),\alpha})$}
\STATE Set $\mathbf{a}^{(n)}_{\beta}$ and the index of chunks in $\mathbf{w}_{v_n^*}^{(n),\beta}$ same as the ones in $\mathbf{w}_{v_m^*}^{(m),\alpha}$.
\STATE Set the order of messages in $\mathbf{w}_{v_n^*}^{(n),\beta}$ the same as the order of messages in $\mathbf{w}_{v_m^*}^{(m),\alpha}$.
\STATE $\gamma$ = Index of the desired message in $\mathbf{w}_{v_n^*}^{(n),{\beta}}$.
\STATE $\mathbf{a}^{(n)}_{\beta}(\gamma) = \mathbf{a}^{(n)}_{\beta}(\gamma)\oplus1$.
\ELSE
\STATE Select $\mathbf{a}_{\beta}^{(n)}= ({a}_{{\beta}, 1}^{(n)}, ..., {a}_{{\beta}, K^{N-2}}^{(n)})$ with i.i.d elements and uniform distribution $\{0, 1\}$.
\STATE Assign new indices for the messages in $\mathbf{w}_{v_n^*}^{(n),{\beta}}$.
\ENDIF
\STATE $Q^{\mathbf{v}^*}_n = Q^{\mathbf{v}^*}_n \cup (\mathbf{a}^{(n)}_{\beta}, \mathbf{w}_{v_n^*}^{(n),{\beta}})$.
\STATE $\beta = \beta+1$.
\ENDFOR
\ENDFOR
\STATE Use $Q^{\mathbf{v}^*}_n$ to request from Server $n$.
\ENDFOR
\STATE Using received answers ($A^{\mathbf{v}^*}_n$, $n\in[N]$), Compute $W^{\mathbf{v}^*}$.
\end{algorithmic}
\end{algorithm}
\begin{algorithm}[tb]
\caption{Server Side Algorithm of an $(N, K)$ DAPAC}
\label{alg_ss_poly}
\begin{algorithmic}[1]
\STATE In Server~$n$: verify attribute $n$ of the user, i.e., $v^*_n$.
\STATE Set  $A^{\mathbf{v}^*}_n=\{\}$ and Type$({i})=0$, $\forall i \in U^{(n)}$.
\FOR {$(\mathbf{a}^{(n)}_{\beta}, \mathbf{w}_{v_n^*}^{(n),{\beta}}) \in  Q^{\mathbf{v}^*}_n$}
\IF {{The attribute $n$ of all messages in $\mathbf{w}_{v_n^*}^{(n),{\beta}}$ is $v^*_n$}, and Type$(T(\mathbf{w}_{v_n^*}^{(n),{\beta}}))==0$} 
\STATE $s_i$ = Part of $\mathcal{C}$ assigned to the type of messages in $\mathbf{w}_{v_n^*}^{(n),{\beta}}$. 
\STATE $A^{\mathbf{v}^*}_n = A^{\mathbf{v}^*}_n \cup \mathbf{a}^{(n)}_{\beta}. \mathbf{w}_{v_n^*}^{(n),\beta}+ s_i$.
\STATE Type$(T(\mathbf{w}_{v_n^*}^{(n),{\beta}}))=1$.
\ELSE
\STATE $A^{\mathbf{v}^*}_n$= \{\}.
\STATE Break.
\ENDIF
\ENDFOR
\STATE Send $A^{\mathbf{v}^*}_n$ to the user. 
\end{algorithmic}
\end{algorithm}

To complete the proof of Theorem~\ref{thoerem_1}, in Section~\ref{proof}, we compute the rate of the proposed scheme and prove that the proposed achievable scheme satisfies the access control and privacy constraints in \eqref{correctness}, \eqref{data_sec_com}, and \eqref{poa} for all $\mathbf{v}^*\in \mathcal{V}^N$. 

\section{Proofs}
\label{proof}
\subsection{Proof of Theorem \ref{thoerem_1}}
\label{theorem_1_proof}
In the proposed DAPAC scheme, we split each message into $\frac{N(N-1)}{2}$ equal chunks, and thus each message chunk has length $\frac{2L}{N(N-1)}$ bits. When the user commits $v^*_n$, he gains access to $K^{N-1}$ messages from Server $n$. But we download these messages in the form of message vectors of length $K^{N-2}$ that, in addition to $v^*_n$, have one common attribute between the other $N-1$ attributes (to satisfy the privacy constraint \eqref{poa}). So the number of linear combinations downloaded from Server~$n$ is equal to $\frac{K^{N-1}(N-1)}{K^{N-2}}$, and the total download from Server~$n$ becomes
\begin{align}
D_n=\frac{K^{N-1}(N-1)}{K^{N-2}}.\frac{2L}{N(N-1)}=\frac{2LK}{N}.    
\end{align}
 
Due to the symmetry between the servers in the scheme, the total download from $N$ servers is,
\begin{align}
D_{t}=ND_n=2LK,   
\label{d_t_poly}
\end{align}
 which is used to retrieve a message with length $L$ bits. So the retrieval rate of the scheme is,
 \begin{align}
 R=\frac{L}{2LK}=\frac{1}{2K}.    
 \end{align}

From \eqref{d_t_poly} and \eqref{download_comp}, by noting that each downloaded equation has $\frac{2L}{N(N-1)}$ bits length, the total number of equations downloaded in this scheme which represents the download complexity  is equal to
\begin{align}
DC=\frac{2LK}{\frac{2L}{N(N-1)}}=KN(N-1),
\end{align}
and the download complexity is of $O(KN^2)$.

To complete the achievability proof, it is required to prove that the proposed scheme satisfies the access control and privacy constraints.

1) \textbf{Access Control:}
We prove the correctness and data secrecy constraints, \eqref{correctness} and \eqref{data_sec_com}, respectively, are satisfied.

\textbf{Correctness:} Suppose that the user with attribute vector $\mathbf{v}^*$ requests the chunks of message $W^{\mathbf{v}^*}$ from all servers. In Algorithm~\ref{alg_ini_poly}, we split each message into $\frac{N(N-1)}{2}$ equal chunks, so it is required to download $\frac{N(N-1)}{2}$ different chunks of $W^{\mathbf{v}^*}$. For each $n, m \in [N]$, $n\neq m$, based on the Definition~\ref{def_4}, $\exists k_n, k_m \in [K]: \Tilde{\mathcal{V}}_m(k_m)= v_m^*, \Tilde{\mathcal{V}}_n(k_n)= v_n^*$, so $U^{(n)}(k_m,m)= U^{(m)}(k_n,n)$. Without loss of generality, suppose $m>n$, then in Algorithm~\ref{alg_us_poly}, the condition in line 9, becomes true, and the user downloads two linear combinations of messages with access policies given in $U^{(n)}(k_m,m)$ in Servers $n$ and $m$ with aligned interferences, and the user can subtract these two linear combinations and retrieve one chunk of the desired message. This argument is true for each ${N \choose 2}$ of servers. So the user can retrieve $\frac{N(N-1)}{2}$ different chunks of the desired message $W^{\mathbf{v}^*}$ (based on line 16 of Algorithm~\ref{alg_us_poly}), therefore we have
\begin{align}
H(W^{\mathbf{v^*}}| A_1^{\mathbf{v^*}}, ..., A_N^{\mathbf{v^*}}, (Q_1^{\mathbf{v^*}}, v^*_1), ..., (Q_N^{\mathbf{v^*}}, v^*_N), \mathcal{P})=0,  
\end{align}
 and the correctness of the scheme is guaranteed.

\textbf{Data Secrecy:} 
First, we prove the data secrecy when the user with attribute vector $\mathbf{v}^*$ sends queries for the message with access policy $\mathbf{v}^*$ to all servers. Next, we prove that if the user sends a query for a message with a different access policy even to one of the servers, then he cannot retrieve the message $W^{\mathbf{v}^*}$ correctly.

Without loss of generality, suppose that the indices of common randomness variables used for the types of messages that comprise $W^{\mathbf{v}^*}$, are in the range of $[\frac{N(N-1)}{2}]$; So, we have
\begin{align}
&I(\mathcal{W}\backslash W^{\mathbf{v}^*}; A_1^{\mathbf{v}^*}, ..., A_N^{\mathbf{v}^*}, (Q_1^{\mathbf{v}^*}, v^*_1), ..., (Q_N^{\mathbf{v}^*}, v^*_N), \mathcal{P}|{W}^{\mathbf{v}^*})\nonumber\\
&=I(\mathcal{W}\backslash W^{\mathbf{v}^*}; \mathbf{a}^{(1)}_1.\mathbf{w}_{v_1^*}^{(1),1}+s_1,..., \mathbf{a}^{(N)}_{N-1}.\mathbf{w}_{v_N^*}^{(N),N-1}+s_{\frac{N(N-1)}{2}}, ..., \mathbf{a}^{(N)}_{K(N-1)}.\mathbf{w}_{v_N^*}^{(N),K(N-1)}\nonumber\\
&+s_{N(N-1)(K-\frac{1}{2})}, (Q_1^{\mathbf{v}^*}, v^*_1), ..., (Q_N^{\mathbf{v}^*}, v^*_N), \mathcal{P}|{W}^{\mathbf{v}^*})=0,\label{data_sec_poly}
\end{align}
where \eqref{data_sec_poly} follows from the independence of common randomness, messages, and queries. So the user gains no information about the messages with different access policies.

Now, we show that if the user sends queries for a message rather than $W^{\mathbf{v}^*}$, then the correctness constraint is violated. Without loss of generality, suppose that the user with attribute vector $\mathbf{v}^*$ sends a query for message $\mathbf{\bar{v}}$ to server $N$. We consider two possible cases below:

(i) Except the attribute $v_N^*$, two vectors $\mathbf{\bar{v}}$ and $\mathbf{v^*}$ have no other attribute in common. Then in server $N$, the coefficients used for the vectors of messages that include the message with access policy $\mathbf{v^*}$ are random, and the user cannot use these equations to retrieve useful chunks of the message $W^{\mathbf{v}^*}$.

(ii) In addition to the attribute $v_N^*$, two vectors $\mathbf{\bar{v}}$ and $\mathbf{v^*}$ have at least one common attribute. Then linear combinations that include messages with access policies $\mathbf{v^*}$ and $\mathbf{\bar{v}}$, cannot be used to retrieve a useful chunk of the message $W^{\mathbf{v}^*}$. Because the coefficients of these linear combinations are tailored to the retrieval of message $W^{\mathbf{\bar{v}}}$.

From the above two cases, we conclude that if the user with attribute vector $\mathbf{v}^*$ sends a query for a message with access policy $\mathbf{\bar{v}}$, then the user cannot retrieve $W^{\mathbf{v}^*}$, completely. This completes the proof of access control constraints.

2) \textbf{Privacy:}
From \eqref{poa}, to preserve the privacy of the other attributes of the user, it is required that $\forall n \in [N]$:
\begin{align}
I(\{v^*_i:i\in [N],i\neq n\};Q_n^{\mathbf{v}}|v^*_n,\mathcal{C},\mathcal{W})=0,
\end{align}
for $\mathbf{v}=(v_1, ..., v_n^*, ..., v_N)$. In the achievable scheme, there are three features below:

(i) The structure of queries from Server $n$ is the same for all users with attribute $v_n^*$. In fact, all users with attribute $v_n^*$ should download $K(N-1)$ linear combinations of messages with access policies given in $U^{(n)}(k,j)$, for each $k \in [K]$ and $j \in [N]\setminus n$. 

(ii) The index of messages are specified after using a uniform and random permutation $\mathcal{P}$ on the indices. 

(iii) For each $\beta \in [K(N-1)]$, each element of the coefficients vectors $a_{\beta}^{(n)}$ has an independent uniform distribution in $\{0, 1\}$.

Therefore, as the structure of the equations is fixed, and the coefficients vector and indices have uniform distributions, each server learns nothing about the attributes of the user, except the one exposed to it, and the privacy of the other attributes of the user is preserved.

\subsection{Proof of Theorem \ref{thoerem_2}}
\label{theorem_2_proof}
In the proposed DAPAC scheme, a part of common randomness $\mathcal{C}$ (with length $\frac{L}{\frac{N(N-1)}{2}}$ bits) is assigned to each set of messages with $K^{N-2}$ chunks of messages that have different access policies and two common attributes. The number of sets of messages with cardinality $K^{N-2}$ composing of messages with different access policies and two attributes in common is,
\begin{align}
{N \choose 2}K^2=\frac{N(N-1)K^2}{2}.
\end{align}
 Since we use independent randomness for different types of messages, then we have a lower bound on the amount of common randomness $\mathcal{C}$ as,
\begin{align}
H(\mathcal{C})\geq\frac{N(N-1)K^2}{2}.\frac{L}{\frac{N(N-1)}{2}}=K^2L,     
\end{align}
which completes the proof of Theorem~\ref{thoerem_2}.

\bibliographystyle{ieeetr}
\bibliography{references}
\end{document}